# A new architecture for making highly scalable applications[1].

**Author: Harry Fitié**

**AB-Ontwikkeling BV**
**Lingedijk 21 4002XH Tiel, The Netherlands**

info@communsens.com

Die Grenze meiner Sprache bedeuten die Grenze meiner Welt [1].
L. Wittgenstein

## Abstract

An application is a logical image of the world on a computer. A scalable application is an application that allows one to update that logical image at **run time**. To put it in operational terms: an application is scalable if a client can change between time T1 and time T2

1. the logic of the application as expressed by language L;
2. the structure and volume of the stored knowledge;
3. the user interface of the application,

while clients working with the application at time T1 will work with the changed application at time T2 without performing any special action between T1 and T2.

In order to realize such a scalable application a new architecture has been developed that fully orbits around language. In order to verify the soundness of that architecture a program has been build. Both architecture and program are called CommunSENS.
The main purpose of this paper is
1. to list the relevant elements of the architecture;
2. to give a visual presentation of how the program and its image of the world look like;
3. to give a visual presentation of how the image can be updated.
Some relevant philosophical and practical backgrounds are included in the appendixes.

## 1 Panta rhei

*Panta rhei*, all flows. These words express the observation that the world is subject to permanent change. In order to live in such a world every creature must have some image of that world. In order to survive in such a world every creature must have some ability to update that image. In a static world it is enough to *be*. In a dynamic world one has to *become*[2]. This becoming reflects the capacity to adapt.

## 2 Scalable applications

An application is a logical image of the world (as perceived by a human being) on a computer. A scalable application is an application that allows you to update that logical image at **run time**. To put it in operational terms: an application is scalable if a client can change between time T1 and time T2

1. the logic of the application as expressed by language L;
2. the structure and volume of the stored knowledge;
3. the user interface of the application;

while clients working with the application at time T1 will work with the changed application at time T2 without performing any special action between T1 and T2.

Suppose there is such a scalable application. Then the initial image of the world that the program contains (say I) can be changed at run time. So if the application to be build should give another image of the world (say X) one should update I to X. In this way a scalable application is not an instant computer program but rather a *problication*: a **pro**gram that **b**ecomes the app**lication**. This becoming never stops: whenever X ceases to be an adequate image of the world, it can be updated to a more proper image.

## 3 Idea

CommunSENS has been based on the proposition of L. Wittgenstein as presented at the title page: *The limits of my language mean the limits of my world*. If that language is 'fixed' the world it can describe is also fixed. On the other hand if the language is not fixed (scalable) it will have the capacity to describe a world that is subject to permanent change.

Figure 1 visualizes the idea. If the language is scalable it can be **upsized** (both in quantity and quality). In general, an expanded language expresses a more 'complete' world. Theoretically, there is no limit to the expansion of a scalable language. If a language is scalable it also can be **downsized**. One can downsize $L_2$ to $L_1$ and $L_1$ to $L_0$. However, in the case of downsizing there are two limits. The first one is an absolute limit. One can not downsize the language beyond the origin O (an empty language). Now the question rises: when does a language becomes a language? Clearly, an empty language does not deserve that name. What about a language that contains one word? What about a language containing 10 words? What about a language containing some words and some syntax? Somewhere between an empty language and a fully developed language there is a point where 'there is language'. The only way to find the **language point** is to use language itself: the language point is that point of the continuum at which the language just defines itself ($L_0$). So the second limit with respect to downsizing is the language point $L_0$[3]. This is the logical limit.

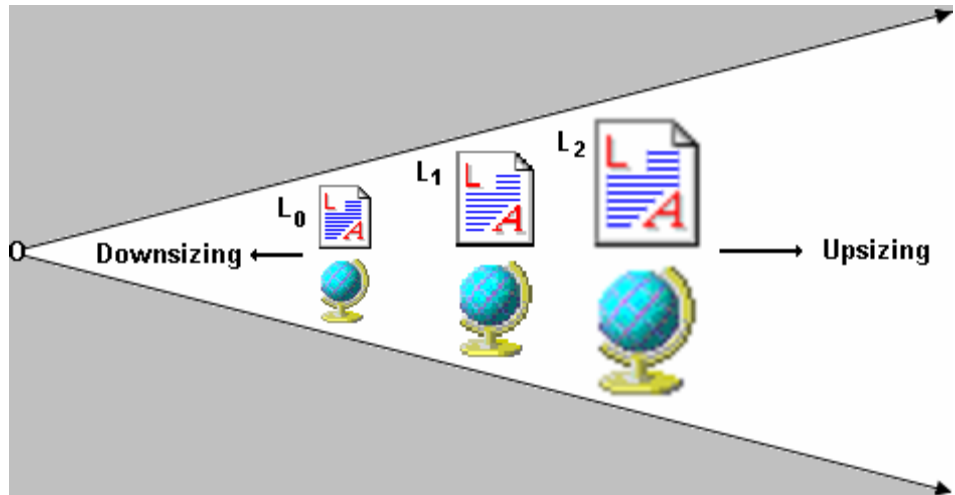

*Figure 1: The limits of my language mean the limits of my world*

## 4 Architecture

Starting with the language, the elaboration of idea to architecture takes eight steps:

1. Define a language that defines itself ($L_0$). This can be compared with writing a book *about* language L *in* language L.
2. Make the language (upsize) scalable with respect to lexicon and syntax. Now, any book can be written in language L.
3. Design a universal database that symbolically stores any utterance expressed in language L.
4. Write the book about the user interface. A communication cycle between man and machine can be started.
5. Design functionality for every kind of *executable* utterance in such a way that it can be applied to the data-agent that applies it.
6. Design a synchronisation mechanism that handles the situation in which executable utterances are modified while they are executing.
7. Define the system in language L and build it in such a way that it could (re-)build itself. The result is a basic image of the world (I).
8. Build an application X by updating image I to image X.

Note that the resulting image of the world is fully closed: **all** information is expressed by language utterances (including user data). Also note that because of $L_0$, the downsize capacity of the architecture has a logical limit.

## 5 Main concepts

Most of the elements of the architecture are explained in some detail in the appendixes. At this place the results of the architecture as a whole will be discussed. Besides language, four main concepts play an important role in the elaboration:

- the concept of self definition;
- the concept of self building;
- the concept of a basic image of the world;
- the concept of updating that basic image.

The first three concepts can be visualized by lithography of the Dutch artist Maurits Escher [2]. Figure 2 shows a picture of two hands that draw each other. In a way, the left hand defines the right hand and the right hand defines the left hand (self definition). The picture has been designed in such a way that it could have been drawn by the hands themselves (self building). The lithography as a whole can be viewed as a basic image. The image is basic because one can not remove essential parts of it (hand nor pencil) without destroying the very idea behind the image. It is holistic.

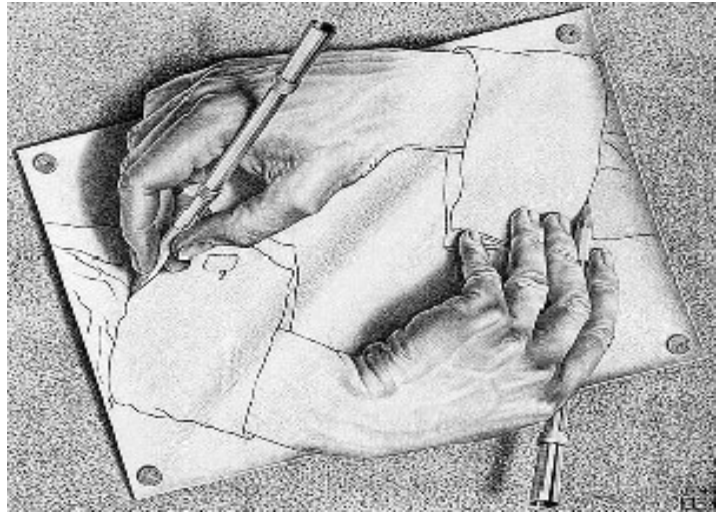

*Figure 2: Maurits Cornelis Escher, 'Tekenende handen', 1948, lithografie.*

## 6 Basic images

Like the hands that have been drawn to draw themselves, CommunSENS has been build to build itself. This resulted in five basic images.

1. $L_0$: the language that defines itself and all other parts of the system;
2. $L_p$: the parser that parses itself and all other parts of the system;
3. $L_d$: the designer that designs itself and all other parts of the system;
4. $L_h$: the help function that helps itself and all other parts of the system;
5. $L_b$: the debugger that partially debugs itself and, fully, all other parts of the system.

The first image of the list is $L_0$. The other images are expansions of that basic image. Although they are 'less basic' they still are holistic. To make this point clear, consider a brain surgeon who performs surgery on his own brain. As long as the result of his surgery does not interfere with his abilities to operate he can continue. However, if the operation affects, for example, his locomotion, the system of the person as surgeon will halt. In general, any scalable system that has functionality that can be applied to change itself is holistic. It can not be downsized beyond a certain point.

As an example, the design form that designs itself, is discussed. Figure 3 shows the main form that is used for designing forms. It consists of two sub forms: FForm at the left side and FDesign at the right side. If a form is selected in FForm its complete visual, logical and behavioural definition is visualized in FDesign. You can update that definition by adding or deleting components, by changing properties of the components, by changing the behaviour of components (events), etc. If the form in question is on the screen, these changes will be immediately reflected.

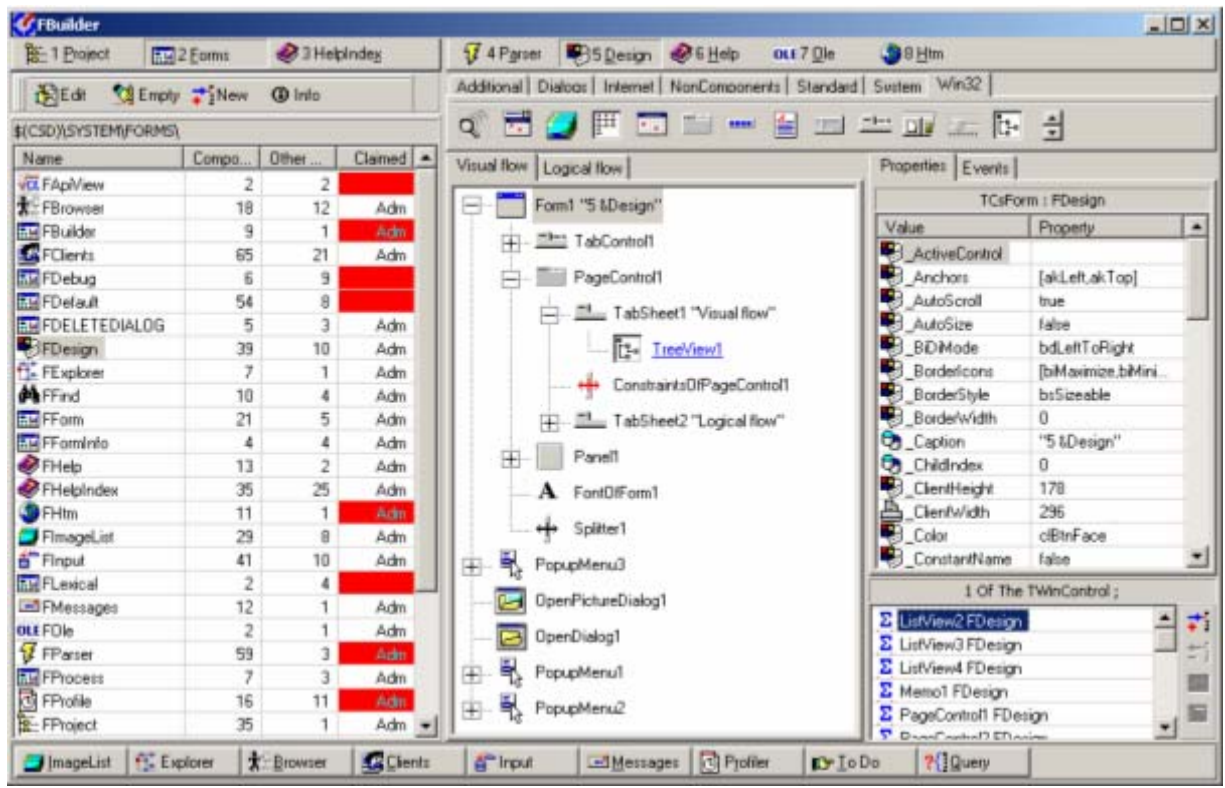

*Figure 3: The design form that designs itself.*

If you look closer at figure 3, you will see that the selected form in FForm (left side) is FDesign. So the right side (FDesign) shows the complete definition of FDesign. In this way FDesign is contained in itself. This feature makes it possible to use FDesign to further design itself. For example, if you add a component from the toolbar to the tree, FDesign will directly show that component. On the other hand, if you remove the component *Tabsheet1 'Visual flow'* both tab sheet and tree will vanish from the form. It is clear that **downsizing** these kinds of basic images should be done with care.

## 7 Updating the basic image

Updating the basic image of the worlds boils down to: updating the language utterances that express that world. Figure 4 shows the main form that has been designed for this purpose. At the left side there is a project tree that allows navigation through all projects. These projects include a linguistic definition of the language itself, linguistic definitions of the builder forms (including the project and parser forms themselves), linguistic definitions of the user interface, etc. At the right side of the main form there is the parse editor. The editor shows the contents of the file that is selected in the project tree. In this editor you can enter new language utterances that define an extended image. Now, updating the initial image is nothing more than parsing (or deleting) these language utterances. If you parse new utterances you act as moulder who uses modelling clay to build the image. If you delete utterances you act as the sculptor who removes those stone particles that hide the image.

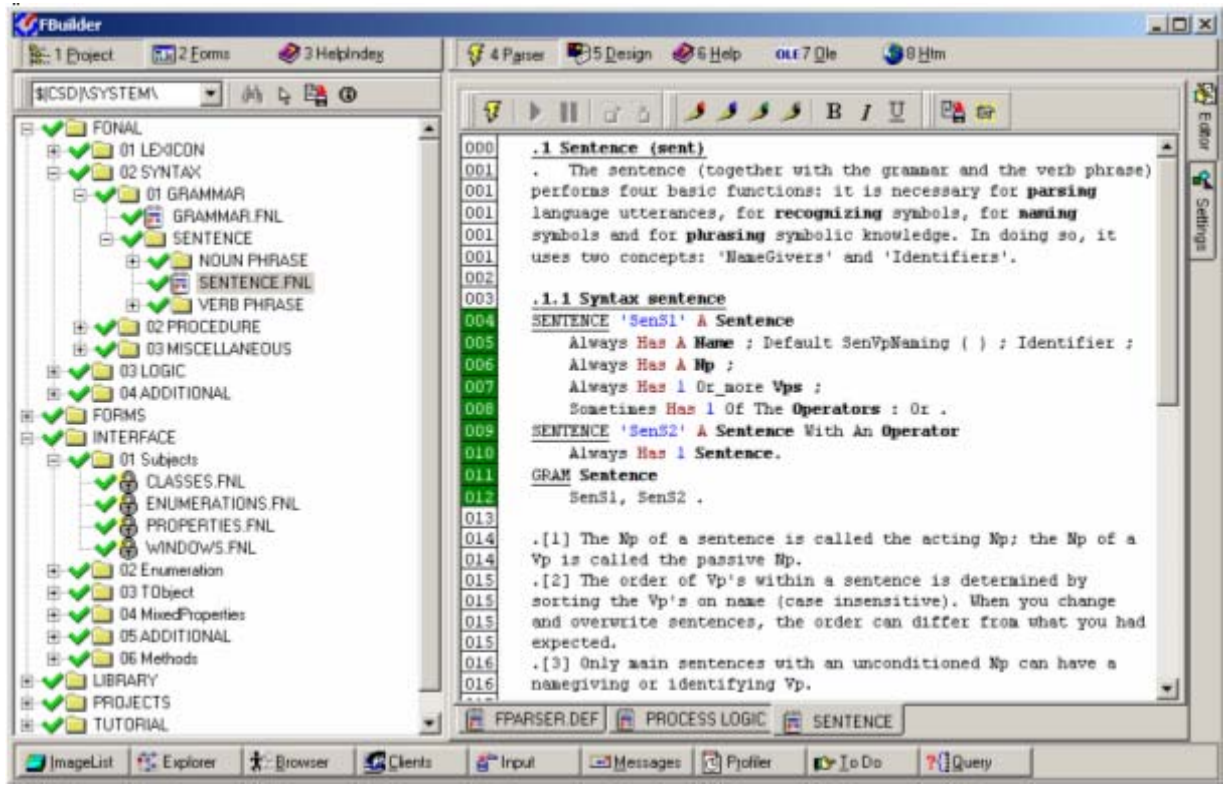

*Figure4: Tools for updating the image of the world*

It should be noted that all changes to the image including those originated from operations as cut, copy, paste, etc. are always expressed as new, updated or deleted language utterances. Therefore, designing a form in FDesign by using the mouse will give the same results as updating the corresponding language utterances in the parser.

## 8 Unity of functionality and knowledge

This simple linguistic update mechanism of the image reveals another feature of the system that is quiet attractive. Normally if one uses tools to build an application, neither the functionality nor the knowledge (F&K) that is contained in the tools will be contained in the new application. Because CommunSENS is an integral part of the new application both are automatically transferred. This works to both sides: old F&K can be applied to an expanded image, new F&K can be applied to the original image. So during the never ending effort of updating the image of the world, there will be unity of knowledge and functionality. Consequently, if the steps mentioned in paragraph 4 are applied to a computer operation system as a whole, unity of knowledge and functionality will be achieved throughout the computer.

## 9 System and restrictions

This paper is concluded with two notes about system and restrictions. The heart of the system is a DLL of about 0.6 Mb. It contains the basic functionality and logic. The basic images, including the language and the definition of the system itself, are stored in the universal database. The basic images can be used as starting point for an update to another image. It also is possible to rebuild the system at run time with these restrictions in mind:

- do not remove any data that is needed for the self definition of the language;
- do not remove any data that gives a single entree to certain functionality.

In terms of the lithography, these restrictions boil down to: do not erase the drawing pencils or the drawing hands.

## Appendix 1 The human paradox

Without doubt, the species that has the most complete image of the world (and even of the 'world' outside his immanent world) is mankind. In this sense, man is the most complete creation.

Why is man the most complete creation? What is the most characteristic other feature of man compared to other creatures? The answer is quite paradoxical: *man is the most complete creation because he is born the least complete* [3].

To make this point clear, enter the world of a 'lower' animal like an ant. An ant is born as an ant. In the process of 'becoming an ant' all essential steps are taken in the foetal period. Once out of the egg, the process of 'becoming an ant' stops. As an ant it directly steps into the ants' world. This world is closed as far as its possibilities are concerned, as it was programmed by the constitution of the creature at birth. Every stimulus from the environment or from the body is handled by instinctive rules or is ignored. In this sense, an ants' world is an absolutely orderly world, without surprises. But it also is a small world

When a human child is born it is not capable of much. While the ant prepares to go to work, the behaviour of the child is very limited. Compared to other creatures from the animal world a human has a limited and unspecialized pattern of instincts: he is remarkably incomplete. As a species, mankind would not survive if this omission was not compensated by the ability of learning and thinking. In contrast to the ant who is born as an ant, a human child is not born as a man but as a 'creature that becomes a man'. Some of the features of 'being a man' are formed in the prenatal period especially the physiological features. Other features like speech, counting, historical consciousness, anticipation, etc. are learned in the years after birth. By learning the relevant rules and how to apply them, a maturing child slowly acquires the world of man that was not given him at birth.

The process of 'becoming a man' does not stop after learning and applying a certain set of rules. After all, where do the learned rules themselves come from? Certain rules can be 'detected by accident' (as a connection between a certain action and a certain result within a certain environment). However most of the rules we learn in our youth have been deliberately fabricated by wise men in past and present. Therefore, the process of 'becoming a man' will never stop as long as there are other men who fabricate new rules worth learning. In this sense, mankind as a whole is involved in a never ending enterprise: creating the world of man.

This line of reasoning has been applied to CommunSENS. When one tries to build a scalable application, the aim should not be to build a program that is complete. After all the result will look like an ant: a highly specialized but rather inflexible program. It is better to build an incomplete program that has the potential of becoming complete. Such a program can be called a Problication (acronym of PROgram Becoming the appLICATION). CommunSENS can be considered as an attempt to build such a problication,

## Appendix 2 Communication cycle

In appendix 1 it was concluded that to become a man a child has to *learn* in order to acquire an image of the world of men. Basically, the process of (declarative) learning takes place in interaction with other people by means of a communication cycle. Figure 5 shows the communication between two men. Person A has knowledge that is stored in his brain. In order to transfer that knowledge he should phrase the knowledge (put it into words). The result is a language utterance that can be transferred to person B (verbal or written). In order to get the meaning out of the utterance, person B has to parse the utterance. After that the contents of the utterance can be added to his image.

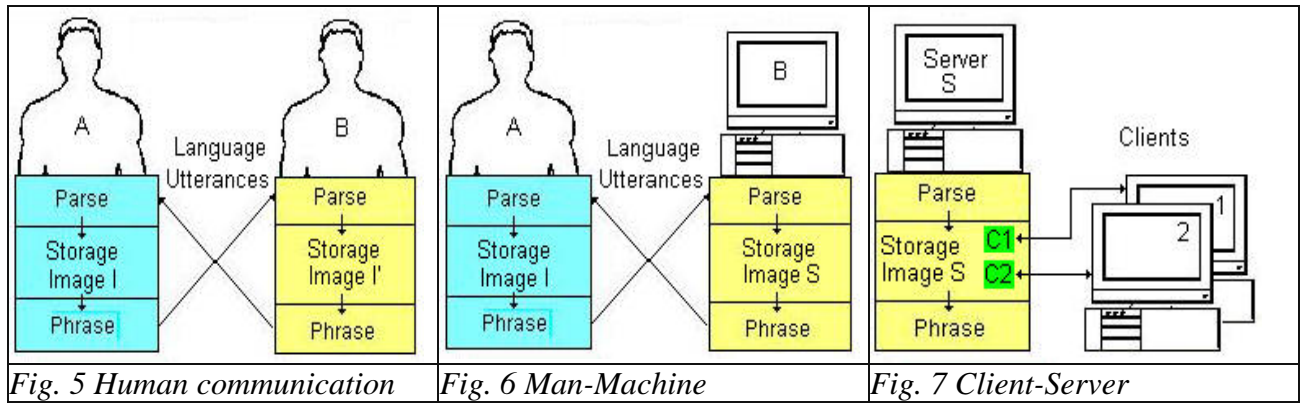

*Fig. 5 Human communication*    *Fig. 6 Man-Machine*    *Fig. 7 Client-Server*

In a similar manner one can design a communication cycle between man and machine (figure 6). If the machine runs a client-server application the architecture will look like figure 7 (with one image for the server application S and one image for each client C that is connected). Note that all information of the client that affects the image goes through the parser (as in figure 6)

CommunSENS is an acronym of COMMUNication between SEmantic Network Systems. In this way the name of the architecture reflects the importance of the communication cycle.

The communication cycle reveals the functional parts that are necessary in order to build an image. One needs

- a **language** in which utterances are formulated
- a mechanism for **recognizing** the symbols that are addressed by the words of the utterance
- a **parser** that transforms the sequence of words into symbolic knowledge.
- a **storage device and structure** to store knowledge about the image.
- a mechanism for **naming** stored symbols;
- a **phraser** to put symbolic knowledge into a sequence of words.
- a **user interface** to give clients a view of the server image.

## Appendix 3 Universal database: SemNET

If one takes a picture of an object the resulting image will resemble that object. In the same way the image of the world will resemble the real world. In his famous work *Tractatus logico-philosophicus* the Austrian philosopher L. Wittgenstein describes this real world by means of two ontological propositions [1]

| Proposition | Description[4] |
|---|---|
| **1** | **The world is everything that is the case.** |
| 1.1 | The world is the totality of facts, not of things. |
| 1.2 | The world divides into facts. |
| **2** | **What is the case, the fact, is the existence of connections.** |
| 2.01 | A connection is a combination of objects (entities, things). |
| 2.011 | It is essential to a thing that it can be a constituent part of a connection. |
| 2.02 | The object is simple. |
| 2.03 | In the connection objects hang one in another, like the links of a chain. |
| 2.04 | The totality of existent connections is the world. |
| 2.05 | The totality of existent connections also determines which connections do not exist. |
| 2.1 | We make ourselves images of facts |
| 2.2 | The image has the logical form of representation in common with what it pictures |

*Table.1: Ontological propositions*

The two features of objects are

- they are distinguishable from each other;
- they can be part of a connection.

What can be gathered from this? First, there is a set of objects or things, say $\Sigma$. Second, a connection can be interpreted as an ordered pair of objects. Third, there is a set, say X, of all ordered pairs of objects. Fourth, the world (W) is an element of the power set of X. Fifth, the image of the world I is an element of the power set of W.

(1) $X = \{(a,b) \mid a \in \Sigma \text{ and } b \in \Sigma\}$

(2) $W \in Pow(X)$

(3) $I \in Pow(W)^5$.

To put it less abstractly, consider X to be a magic block of stone in which each stone particle is connected to each other stone particle. This stone hides countless possible sculptures (a power set). From all the possible sculptures that can be made, a divine sculptor creates one by removing the connections between certain stone particles (W). A human sculptor who just has an ordinary block of stone at his disposal, can only hope to sculpture to his best abilities the most significant part (I) of the the structure (W).

Using the verbal definition of an application that has been given in chapter 2, one can define a server image S as an element of the power set of image I (4). The world of a client C that executes within the server application can be considered as an element of the power set of S (5).

(4) $S \in Pow(I)$.

(5) $C \in Pow(S)$.

By means of his senses man can acquire an image of the world. Because the brain is definitely finite, this image will also be definitely finite. It will look something like this (where a1, a2, b1, .. are symbols that refer to specific objects).

(6) I = {..., (a1, a2), (a1, b1), ...}.

This formula has been elaborated to a semantic network. Basically, the network is defined by a symbol domain, three binary relations and two characteristic functions.

The initial semantic network contains among other things the book about the language, the book about the user interface and the interface itself. With about 150 thousand symbols and 630 thousand ordered pairs the image was shaped. On an average each symbol is connected to 4.1 other symbols.

## Appendix 4 Language Fonal

A **fo**rmal **na**tural **l**anguage (Fonal) has been defined, loosely based on the language universals proposed by the American linguist N. Chomsky [4]. At this place a summary is presented in which the main structure is sketched (see table 2).

Italic text indicates that the category is directly recursive (e.g. InstructionList Has InstructionList). All other phrases and the expression are indirectly recursive (e.g. NP of PP of NP). Syntactic categories in bold, can use some or all of the elements of the last row. A non capitalized category indicates a terminal symbol.

| Syntactical structures | | | | | | | | | | |
|---|---|---|---|---|---|---|---|---|---|---|
| Grammar (G) | | | | | | | | | Procedure (P) | |
| *Sentence (S)* | | | | | | | | | *Instruction list (IL)* | |
| Noun phrase (NP) | | | | | | *Verb phrase (VP)* | | | **Expres-** | *EP* |
| *AP* | **DEFP** | DP | *PP* | | subj | **PNP** | vq | v | NP | **sion** | NP |
| adj | | | *QP* | prep | NP | | | | | **(EXP)** | |
| Numbers, variables, operators, EXP, EP, Procedure calls | | | | | | | | | | |

*Table 2: Summary of syntax of Fonal*

Table 3 gives some examples of language utterances. Most of the examples are borrowed from the book about Fonal in Fonal.

With respect to the image, the purpose of language is four fold

- To prescribe the contents of the image (grammar and sentences)
- To express how one should navigate through the image (prepositional phrases)
- To express how one should query the image (noun phrases)
- To express how one should execute orderly activities on the image (procedures).

| Abbreviation | Syntactical category | \<Bold text\> is an instance of the category |
|---|---|---|
| adj | Adjective | A **Collecting** NP |
| AP | Adjective phrase | **Binary And Relational** Operators |
| DEFP | Default phrase | **Default: {The First Subject}** |
| DP | Determiner phrase | **1 Of The** Subjects |
| EP | Expression phrase | **{All Operators} Minus {Operator: Minus}** |
| EXP | Expression | **A = {Binary Operators} Union {Verbs}** |
| G | Grammar | **Sentence SenS1, SenS2.** |
| IL | Instruction list | 'MyProc'() **{A = 2; Return A;}** . |
| NP | Noun phrase | **A Sentence** Always Has A NP |
| P | Procedure | **'MyProc'() {A = 2; Return A;}** . |
| PNP | Proper name phrase | Subject**: Subject** |
| PP | Prepositional phrase | NP **With Subject: Subject** |
| prep | Preposition | NP **With** Subject: Subject |
| QP | Quantor phrase | **1** Of The Subjects |
| S | Sentence | **'SenS1' A Sentence Always Has A NP** |
| Subj | Subject (Noun) | **Subject** |
| V | Verb | A Sentence Always **Has** A NP |
| VP | Verb phrase | A Sentence **Always Has A NP** |
| Vq | Verb quantor | A Sentence **Always** Has A NP |

*Table3: Examples of utterances of Fonal*

## Appendix 5 User interface

CommunSENS contains a complete, logical definition of the user interface in Fonal. This definition stands apart from the actual implementation of the interface on a computer or operating system. The visual structure of the user interface is based on the taxonomy of Borland's Visual Component Library [5] which in turn rests on MS Windows. Two kinds of properties are distinguished. The first category deals with properties that determine the visual aspects of components as seen by the user. The second category is specifically designed for use with Fonal. These properties, called set events, deal with the way sets are retrieved, updated, changed etc. Table 4 shows the most important events that can be used to conduct the behaviour of components.

The user interface has been designed in such a way that any set, connected to a component by the OnGetSet event, can be used as the context for other sets. For example, if component C1 shows the result of the noun phrase *All subjects*, a second component C2 can show the result of the noun phrase *All symbols classified by **this** subject*. If you select in C1 the subject *Patients* you will get in C2 all stored patients. If you select in C1 the subject *Procedures*, you will get in C2 all stored procedures. The only thing to be done is setting a property that says that C1 is the context for C2. In this way most programming will reduce to an enumeration of noun phrases that are mutually connected by context properties. The forms of figure 3 and 4 have been designed in this way.

| Set events | Generated at | Description |
|---|---|---|
| OnGetSet | Server | Executes a noun phrase that retrieves the symbol set. E.g.. 'All patients' |
| OnGetChildren | Server | Executes a noun phrase that retrieves the node children for tree-like components |
| OnGetParents | Server | Executes a noun phrase that retrieves the node parents for tree-like components |
| OnSetChanged | Server | Executes a free expression when the set changes (symbols added or deleted) |
| OnSelChange | Client | Executes a free expression when the selection changes. Moreover, it automatically updates the symbol stream for context sensitive components |
| OnGetName | Server | Executes a procedure to get the name of a symbol |
| OnGetColumnName | Server | Executes a procedure to get column information |
| OnGetState | Server | Executes a free expression to get the perfect / imperfect state of a symbol |
| OnSelUpdate | Server | Executes a free expression when a selected symbol has been updated |
| OnCommit | Server | Executes 1 or more noun phrases to collect symbols that are affected by an image update |
| OnClick | Client | Executes a free expression |
| OnDblClick | Client | Executes a free expression |

*Table 4 Most important events*

## Appendix 6 Client-server architecture

The server has the important task to keep track of the logical and visual state of all information that is supplied by a connected client. The knowledge state is represented by the client image within the server image (see figure 7).

Communication between server and client is realized by messages that contain strings and symbolic information. For any client that logs in, the server creates a thread. Each thread performs three tasks:

- It decodes messages from the client. These messages contain simple information like *Symbol #1000 is selected in component #1050* or *Component #1100 is clicked.*
- It executes the event (say #1200) that is connected to the specific component.
- It stores resulting information and returns the information as strings and symbols. At the client side this information is made suitable for further visual processing.

For theoretical reasons, at any point in time, there is only one thread (the committing thread) that can update the image. During that time, all other threads are suspended. Because CommunSENS is highly scalable it is possible that while thread T1 is suspended in a procedure, a committing thread T2 deletes this very procedure. It is obvious that when T1 resumes execution, there is a problem if precautions are not taken.

During execution, every thread keeps track of its **execution path**. This path is just a stack of symbols as shown in table 5.

| Executable | Symbol |
|---|---|
| ... | ... |
| Instruction list | #4002 |
| Procedure | #4000 |
| Event | #1200 |
| Listview | #6532 |
| Form | #1326 |
| Client | #1101 |
| Main Entry | Start |

*Table 5 Execution path of a thread*

The stack is pushed on entry of the executable and popped on exit. Now when a committing thread has updated the image it checks the execution path of every suspended thread to see if it uses changed information. If so, it sets a crawl back flag. When committing is ready, the suspended threads resume execution. If the flag is set, they crawl back to the main entree point and try again. If the flag is not set they continue execution. If the execution can not continue (e.g. because a complete component like form #1326 has been removed) the user will be informed that his action can not be performed.

---

[1] This research has been made possible by a grant of the Ministry of Economic Affairs of the Netherlands. SO/2003/11629/1/2565.

[2] In this respect the question is not: to be or not to be. One could say: to become or not to become, *that* is the question.
[3] Note that the language point $L_0$ is relative, depending on the richness of the language. To my knowledge, the smallest complete language is defined by Backus-Naur (as BNF-rules). The most extensive complete language is natural language.
[4] Following the Dutch translation of W.F. Hermans (Amsterdam, 1989) the word 'connection' is used instead of 'atomic fact'. The word 'picture' has been replaced by 'image'.
[5] The domain of objects is not the same for each image. For the human image the domain will be something like the neurons in the brain with the axons representing the connections. When the image is on a machine, the domain of objects will be numbers stored in memory positions with pointers to these positions representing the connections.